\documentclass{elsarticle}
\usepackage[utf8]{inputenc}
\usepackage{graphicx}
\usepackage{amsfonts}
\usepackage{amssymb}
\usepackage{amsmath}
\usepackage{txfonts}
\usepackage{lipsum}
\usepackage{color}
\usepackage{wasysym}
\usepackage{hyperref}
\usepackage{bbold}
\usepackage{etoolbox}
\usepackage{feynmp-auto}

\newcommand{\kv}{\mathbf{k}}
\begin{document}
\begin{frontmatter}
\title{Dual fermion method as a prototype of generic reference-system approach for correlated fermions}

\author[UHH,CUI]{Sergey Brener}
\ead{sbrener@physnet.uni-hamburg.de}
\author[UHH]{Evgeny A. Stepanov}
\author[Skol]{Alexey N. Rubtsov}
\author[Nij]{Mikhail I. Katsnelson}
\author[UHH,CUI]{Alexander I. Lichtenstein}

\address[UHH]{I. Institute of Theoretical Physics, University of Hamburg, Jungiusstrasse 9, 20355 Hamburg, Germany}
\address[CUI]{The Hamburg Centre for Ultrafast Imaging, Luruper Chaussee 149, 22761 Hamburg, Germany}
\address[Skol]{Russian Quantum Center, Skolkovo innovation city, 121205 Moscow, Russia}
\address[Nij]{Radboud University, Institute for Molecules and Materials, 6525AJ Nijmegen, The Netherlands}

\date{\today}

\begin{abstract}
    We present a purely diagrammatic derivation of the dual fermion scheme [Phys. Rev. B 77 (2008) 033101]. The derivation makes particularly clear that a similar scheme can be developed for an arbitrary reference system provided it has the same interaction term as the original system. Thereby no restrictions are imposed by the locality of the reference problem or by the nature of the original problem as a lattice one. We present new arguments in favour of keeping the dual denominator in the expression for the lattice self-energy independently of the truncation of the dual interaction. As an example we present the computational results for the half-filled 2D Hubbard model with the choice of a $2\times2$ plaquette with periodic boundary conditions as a reference system. We observe that obtained results are in a good agreement with numerically exact lattice quantum Monte Carlo data.
\end{abstract}

\end{frontmatter}

\begin{fmffile}{diagram}

\section{Introduction}

The dynamical mean-field theory (DMFT)~\cite{DMFT} has opened new ways in theory of correlated electron systems, in particular, due to its implementation into electronic structure calculations and the related progress in description and understanding the properties of real materials~\cite{Anisimov_1997, Lichtenstein_Katsnelson_1998, Kotliar_RMP_1996}. 
It originates from the consideration of a rather artificial limit of infinite dimensionality~\cite{Metzner_Vollhardt_1989} and strictly speaking is exact only in this limit. Viewing the DMFT approximation in terms of the Luttinger-Ward functional~\cite{Kotliar_RMP_1996} sheds light on the question why it works surprisingly well for real systems that are three- or sometimes even two-dimensional.
The dual fermion (DF) approach~\cite{DF} gave a different view on the DMFT. 
In this approach a change of variables in the path integral over fermionic degrees of freedom is suggested (which can be considered as a functional analog of Fourier transformation) such that in the new variables DMFT Green's function is just the bare Green's function, and a regular diagrammatic expansion starting from this new zeroth-order approximation is possible. 
Specific applications of DF as well as other diagrammatic approaches beyond DMFT are reviewed in Ref.~\cite{Rohringer_RMP_2018}. 
Here we rederive DF theory by a completely different approach, using topological analysis of Feynman diagrams instead of explicit manipulations with the functional integrals. 
The advantage of this view is that it allowed for new insights and possible nontrivial generalizations.
We consider DF as a particular case of the reference system ideology where one relates the initial system described by the Hamiltonian $H$ to some auxiliary system, easier to treat, described by the Hamiltonian $H^*$ as was initially proposed in Peierls-Feynman-Bogoliubov variational principle~\cite{Peierls_1938, Feynman_1955, Bogolyubov_1958} (for recent applications of this method to correlated fermions see~\cite{Schuler_2013, vanLoon_2016}). 
The problem with variational approaches is that generally speaking there is no regular way to improve them systematically. 
Instead, we develop here a diagrammatic approach to the reference system, with two types of Green's functions, related to the system with the Hamiltonian $H$ and to the system with the Hamiltonian $H^*$. 
We show that DF can be considered as a particular case of this approach when $H^*$ corresponds to the effective impurity \emph{i.e.} a lattice site plus a bath~\cite{DMFT}. We consider here a generalization of this approach. As a simple numerical test we use the Hubbard model on the square lattice as an example, with a plaquette playing the role of the reference system~\cite{CDMFT}. 
The results look promising. 
The approach developed here can have more applications, for example, it may potentially help to solve the famous sign problem in quantum Monte Carlo (QMC) calculations~\cite{Loh_1990}, by mapping of the system with sign problem (e.g., $t-t'$ Hubbard model on a bipartite lattice with finite doping) onto the model without sign problem (e.g., the same problem with $t'$ and doping being equal to zero).  

The paper is organized as follows: in Section 2 we give a brief review of the DF formalism and outline some concepts used further. In Section 3 we give a detailed derivation of the generalized DF from the diagrammatic point view. Section 4 is devoted to a discussion of the truncation of the DF scheme. In Section 5 we show numerical results for a simple test problem and compare it with diagrammatic-QMC results~\cite{DiagDFQMC}. Finally in Section 6 we give conclusions and a brief outlook.

\section{Dual fermions}

The dual fermion technique~\cite{DF} was primarily developed as a tool to go beyond the well-established DMFT approximation~\cite{DMFT}. 
It allows to systematically calculate non-local corrections to the (by definition) local self-energy of DMFT. 
The approach was derived using a Hubbard-Stratonovich transformation of the non-local part of the action. 
It is done by separating the action of the Hubbard model
\begin{equation}
{\cal S} = -\sum_{\kv,\nu,\sigma} c^{*}_{\kv\nu\sigma} \left[i\nu+\mu-\varepsilon^{\phantom{\dagger}}_{\kv}\right]
\, c^{\phantom{*}}_{\kv\nu\sigma}+U\sum_i n_{i\uparrow}n_{i\downarrow},
\label{eq:S_latt}
\end{equation}
into the impurity part ${\cal S}_{\mathrm{imp}}$ and the remainder ${\cal S}_{\mathrm{rem}}$:
\begin{align}
{\cal S}_{\mathrm{imp}}=&-\sum_{i,\nu,\sigma} c^{*}_{i\nu\sigma} \left[i\nu+\mu-\Delta^{\phantom{\dagger}}_{\nu}\right]
\, c^{\phantom{*}}_{i\nu\sigma}+U\sum_i n_{i\uparrow}n_{i\downarrow}\\
{\cal S}_{\mathrm{rem}}=&\sum_{\kv,\nu,\sigma} c^{*}_{\kv\nu\sigma} \left[\varepsilon^{\phantom{\dagger}}_{\kv} - \Delta^{\phantom{\dagger}}_{\nu}\right]
\, c^{\phantom{*}}_{\kv\nu\sigma}.
\label{S_rem}
\end{align}
Here $c,c^*$ are the fermionic Grassmanian fields, $\kv,\nu,$ and $\sigma$ are the fermion's wave vector, Matsubara frequency, and spin respectively. $\varepsilon_{\kv}$ is the dispersion of the fermions, $n_{i\sigma}$ is the number of fermions on site $i$ with spin $\sigma$, $U$ is the Hubbard interaction constant, and $\mu$ --- the chemical potential.
In Eq.~\eqref{S_rem} we explicitly use that the summation of the hybridization term $c^*\Delta c$ over the wave-vector $\kv$ is identical to summation over the lattice sites $i$ as $\Delta$ is site-independent. 
After this the remainder term in the partition sum is decoupled using Hubbard--Stratonovich transformation introducing the dual fermions $f,f^*$ thus leaving the original fermionic fields $c,c^*$ only in the impurity part, which allows to integrate them out assuming the solution of the impurity problem ${\cal S}_{\mathrm{imp}}$ is known exactly. 
As a result we end up with the dual action
\begin{equation}
{\cal S}_{\mathrm d} = -\sum_{\kv,\nu,\sigma} f^{*}_{\kv\nu\sigma} \tilde{G}^{-1}_0(\kv,\nu) f^{\phantom{*}}_{\kv\nu\sigma}+V[f,f^*].
\label{eq:S_dual}
\end{equation}
Here 
\begin{equation}
\tilde{G}_0(\kv,\nu) = \left(g_{\nu}^{-1}+\Delta_{\nu}-\varepsilon_{\kv}\right)^{-1}-g_{\nu}
\label{g_dual_def}
\end{equation}
is the bare dual Green's function, $g_{\nu}$ being the full impurity Green's function, and
\begin{align}
V[f,f^*] &= \frac14\sum_{\{k,\nu\,\sigma\}}\gamma^{(4)}_{\{\nu,\sigma\}}f_{\{k,\nu\,\sigma\}}^*f_{\{k,\nu\,\sigma\}}^*f_{\{k,\nu\,\sigma\}}f_{\{k,\nu\,\sigma\}}\\&+\frac{1}{36}\sum_{\{k,\nu\,\sigma\}}\gamma^{(6)}_{\{\nu,\sigma\}}f_{\{k,\nu\,\sigma\}}^*f_{\{k,\nu\,\sigma\}}^*f_{\{k,\nu\,\sigma\}}^*f_{\{k,\nu\,\sigma\}}f_{\{k,\nu\,\sigma\}}f_{\{k,\nu\,\sigma\}}+\dots
\label{interaction}
\end{align}
is the interaction of the dual fermions that is represented here symbolically meaning that the complete dual interaction consists from $n$-particle local interactions for all $n>1$, and the role of the $n$-particle interaction function is taken by the full $n$-particle vertex $\gamma^{(2n)}$ of the impurity problem. 
The summation variables $\{k,\nu\,\sigma\}$ represent the wave-vector, the frequency and the spin. 
All usual conservation laws are, of course, fulfilled reducing the number of independent variables in each term to $2n-1$. 
The choice of these parameters depends on the particular problem under consideration and we will not specify it, while for our purposes it is enough to say that the usual diagram building rules apply. 
We will concentrate on the topology and partial summation of diagrams~\cite{AGD} and not on the calculations themselves.

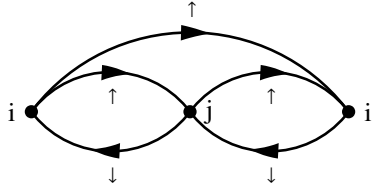
\begin{figure}
\centering
  \begin{fmfgraph*}(120,40)
  \fmfleft{i}
  \fmfright{o}
  \fmf{fermion,label=$\phantom{|}^\uparrow$,label.side=right,left=.5}{i,v}
  \fmf{fermion,label=$\phantom{|}^\downarrow$,left=.5}{v,i}
  \fmf{fermion,label=$\phantom{|}^\downarrow$,left=.5}{o,v}
  \fmf{fermion,label=$\phantom{|}^\uparrow$,label.side=right,left=.5}{v,o}
  \fmf{fermion,label=$\phantom{|}_\uparrow$,left=.5}{i,o}
  \fmfdot{i}
  \fmfdot{o}
  \fmfdot{v}
  \fmflabel{i}{i}
  \fmflabel{i}{o}
  \fmflabel{j}{v}
  \end{fmfgraph*}
    \caption{An example of a local contribution to the self-energy not included in DMFT. If the sites $i$ and $j$ are different, then this diagram is local, but is clearly not included in $\Sigma_{\mathrm{imp}}$.}
    \label{Fig1}
\end{figure}

The usual way of practical implementation of the dual fermion formalism is to truncate the dual interaction restricting oneself only to the 2-particle term, and also to choose the hybridization function $\Delta_{\nu}$ in such a way that the DMFT self-consistency condition is fulfilled, \emph{i.e.}
\begin{equation}
\frac{1}{N_k} \sum_{\kv}G^{\mathrm{DMFT}}_{\kv\nu}=g_{\nu}.
\label{eq:condition}
\end{equation}
Here $G^{\mathrm{DMFT}}_{\kv\nu}=(g_{\nu}^{-1}+\Delta_{\nu}-\varepsilon_{\kv})^{-1}$ is the DMFT Green's function, and $N_k$ is the number of $k$-points. From Eq.~\eqref{g_dual_def} it is straightforward to see that the bare dual Green's function is given by $\tilde G_0=G^{\mathrm{DMFT}}-g$. 
This actually means that the starting point of the dual fermion formalism for this particular choice of the hybridization is the DMFT solution, and any correction to it stemming from the dual fermion diagrammatics is a beyond DMFT correction. 
Even the simplest possible corrections have proven to bring interesting insight in the properties of the Hubbard model, but there are ideological problems with the method mostly due to the lack of clear understanding of what effects are accounted for by what classes of dual diagrams. 
It is still an open question, for example, when and to which extent leaving out the higher-order interaction terms (which are very hard to handle in practical calculations) is justified.\cite{DiagDFQMC,Hafermann_2009_2,Ribic_2017,Ribic_2017_2,Katanin_2013} 
The whole dual technique is not very transparent due to the non-trivial nature of the Hubbard--Stratonovich transformation that is used to introduce dual fields. 
The common wisdom is that the dual Green's function is somehow connected to the non-local part of the original Green's function, but this connection is literal only for the DMFT based choice of the hybridization and only for the zeroth order. 
In any other case this is strictly speaking not true.
There exists a well-known exact connection between the original $G$ and dual $\tilde G$ Green's functions~\cite{DF}:
\begin{equation}
G=(\Delta-\varepsilon)^{-1}+(\Delta-\varepsilon)^{-1}g^{-1}\tilde G g^{-1}(\Delta-\varepsilon)^{-1}.
\end{equation}
Here, we omit the indices $\kv,\nu,\sigma$, which are trivially restored, and write the above formula in a way that underlines its symmetry which becomes essential in a multi-orbital case or in the case with a broken symmetry (superconductivity, magnetism), where the matrix nature of the functions above is relevant. 
There is another way of writing down the exact connection between the original and dual fermions, which is more compact and allows more insight. For this we have to address the self-energies instead of the Green's functions. The (exact) connection becomes~\cite{PhysRevB.79.045133}:
\begin{equation}
\Sigma=\Sigma_{\mathrm{imp}}+\tilde\Sigma(1+g\tilde\Sigma)^{-1}.
\label{sigma_prime_dual}
\end{equation}
Here, $\Sigma$ is the self-energy of the original problem, $\Sigma_{\mathrm{imp}}$ is the self-energy of the impurity problem, and $\tilde\Sigma$ is the self-energy of the dual problem. The last term on the r.h.s. is often denoted as $\Sigma'$ and represents the correction to the impurity self-energy. It would be wrong to call $\Sigma'$ the non-local part of the self-energy as it has local contributions. A simple diagram in Fig.~\ref{Fig1} is an example of such a contribution.

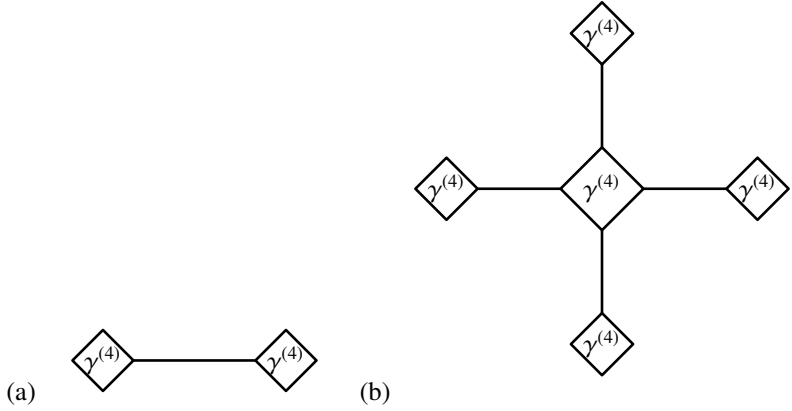
\begin{figure}
    \centering
    \vspace{-3cm}
    (a)\begin{fmfgraph*}(120,30)
    \fmfleft{i}
    \fmfright{o}
    \fmfpolyn{empty,tension=0.3,label=$\gamma^{(4)}$}{K}{4}
    \fmfpolyn{empty,tension=0.3,label=$\gamma^{(4)}$}{L}{4}
    \fmf{phantom}{i,K1}
    \fmf{phantom}{o,L3}
    \fmf{plain,tension=0.3}{K3,L1}
    \end{fmfgraph*}
    (b)\begin{fmfgraph*}(160,160)
    \fmfleft{i}
    \fmfright{o}
    \fmftop{t}
    \fmfbottom{b}
    \fmfpolyn{empty,tension=0.2,label=$\gamma^{(4)}$}{K}{4}
    \fmfpolyn{empty,tension=0.2,label=$\gamma^{(4)}$}{L}{4}
    \fmfpolyn{empty,tension=0.2,label=$\gamma^{(4)}$}{M}{4}
    \fmfpolyn{empty,tension=0.2,label=$\gamma^{(4)}$}{N}{4}
    \fmfpolyn{empty,tension=0.3,label=$\gamma^{(4)}$}{V}{4}
    \fmf{phantom}{i,K1}
    \fmf{phantom}{t,L4}
    \fmf{phantom}{o,M3}
    \fmf{phantom}{b,N2}
    \fmf{plain,tension=0.3}{K3,V1}
    \fmf{plain,tension=0.3}{L2,V4}
    \fmf{plain,tension=0.3}{M1,V3}
    \fmf{plain,tension=0.3}{N4,V2}
    \end{fmfgraph*}
    \caption{Examples of a one-particle reducible contributions to higher-order vertices. Hereafter plain lines represent the full impurity Green's functions $g$. We also omit the arrows in order not to over-complicate the graphs.}
    \label{gamma_reduc}
\end{figure}

The meaning of the denominator $(1+g\tilde\Sigma)$ in the expression for $\Sigma'$ becomes clear if we notice that it represents nothing else then a part of $\tilde\Sigma$ which is irreducible in $g$. 
Indeed, the formula $\Sigma'=\tilde\Sigma(1+g\tilde\Sigma)^{-1}$ is equivalent to $\tilde\Sigma=\Sigma'(1-g\Sigma')^{-1}$ and the above statement becomes trivial. 
To understand what this means requires some insight in the structure of the dual technique~\cite{Katanin_2013}. 
As mentioned above the dual interaction consists from all possible many-particle terms. 
These terms are given by the full $2n$-point vertices of the impurity problem. 
The latter are defined as the non-trivial part of the $n$-particle Green's functions $G^{(n)}$ with amputated external legs. 
For example
\begin{equation}
    \gamma^{(4)}_{1234}=\sum_{1',2',3',4'}(g^{-1})_{11'}(g^{-1})_{33'}\left[G^{(2)}_{1'2'3'4'}-g_{1'2'}g_{3'4'}+g_{1'4'}g_{3'2'}\right](g^{-1})_{2'2}(g^{-1})_{4'4}.
\end{equation}
Here, the numbers represent the combined indices including spin, frequency, and, if applicable, other impurity degrees of freedom. 
The last two terms in the square brackets are the trivial part of the two-particle Green's function that are subtracted from it. 
Higher order vertices are defined similarly. 
It is easy to recognize that this definition forbids contributions to $\gamma^{(4)}$ that are reducible in $g$. But higher order vertices can have contributions that are one-particle reducible. 
The simplest example is a contribution to $\gamma^{(6)}$ represented by two bare interaction vertices $U$ connected by $g$. 
In general, we can connect two $\gamma^{(4)}$ vertices by a single $g$-line and get a one-particle reducible contribution to $\gamma^{(6)}$ (Fig.~\ref{gamma_reduc}\,(a)). 
For higher order vertices the possibilities get even richer. Fig.~\ref{gamma_reduc}\,(b) shows a contribution to $\gamma^{(12)}$ that is one-particle reducible in four different points.

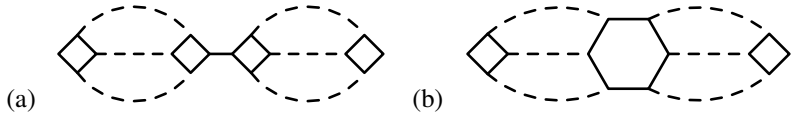
\begin{figure}
    \centering
    (a)\begin{fmfgraph*}(140,40)
    \fmfleft{i}
    \fmfright{o}
    \fmfpolyn{empty,tension=0.3}{K}{4}
    \fmfpolyn{empty,tension=0.3}{L}{4}
    \fmfpolyn{empty,tension=0.3}{M}{4}
    \fmfpolyn{empty,tension=0.3}{N}{4}
    \fmf{phantom}{i,K1}
    \fmf{phantom}{o,L3}
    \fmf{dashes,tension=0.3}{K3,M1}
    \fmf{dashes,tension=0.3}{N3,L1}
    \fmf{plain}{M3,N1}
    \fmffreeze
    \fmf{dashes,left=.5}{K4,M4}
    \fmf{dashes,left=.5}{N4,L4}
    \fmf{dashes,right=.5}{K2,M2}
    \fmf{dashes,right=.5}{N2,L2}
    \end{fmfgraph*}
    (b)\begin{fmfgraph*}(140,40)
    \fmfleft{i}
    \fmfright{o}
    \fmfpolyn{empty,tension=0.3}{K}{4}
    \fmfpolyn{empty,tension=0.3}{L}{4}
    \fmfpolyn{empty,tension=0.2}{M}{6}
    \fmf{phantom}{i,K1}
    \fmf{phantom}{o,L3}
    \fmf{dashes,tension=0.3}{K3,M1}
    \fmf{dashes,tension=0.3}{M4,L1}
    \fmffreeze
    \fmf{dashes,left=.3}{K4,M6}
    \fmf{dashes,left=.3}{M5,L4}
    \fmf{dashes,right=.3}{K2,M2}
    \fmf{dashes,right=.3}{M3,L2}
    \end{fmfgraph*}
    \caption{(a) $g$-reducible contribution to the dual self energy. The two 4-point vertices connected by the impurity line are a part of a 6-point impurity vertex. (b) The corresponding dual diagram with a 6-point vertex. The dual Green's functions are depicted by dashed lines.}
    \label{g_reduc_sigma}
\end{figure}

Although for the dual diagrammatic technique those vertices, whether or not they are one-particle reducible, enter on equal footing, we can still consider all vertices as a sum of the one-particle irreducible part and all possible one-particle reducible parts. 
And that leads us to the understanding of what the irreducible in $g$ part of $\tilde\Sigma$ mentioned above means. 
The simplest example of a $g$-reducible contribution to the dual self-energy is shown on Fig.~\ref{g_reduc_sigma}.

\section{Diagrammatic derivation of the dual technique.}

Up to now we have shown that the correction to the impurity self-energy that gives the full self-energy is nothing else than the $g$-irreducible part of the dual self-energy. 
This has been shown using the exact relation between the above-mentioned quantities and by diagrammatic interpretation of this relation. 
In the following we will examine the diagrams that contribute to $\Sigma'$ and gain a diagrammatic insight of why those diagrams indeed sum up to the $g$-irreducible part of the dual self-energy. 

At this point let us briefly return to the beginning and generalize our consideration. 
From this point on we no longer restrict ourselves to any specific model like the Hubbard one. 
We consider an arbitrary fermion action with the general two-particle (antisymmetrized) interaction $U_{1234}$:
\begin{equation}
    {\cal S}=-\sum_{1,2} c^*_1[G_0^{-1}]_{12}c_2+\frac14\sum_{1,2,3,4}U_{1234}c^*_1c^*_2c_3c_4,
    \label{eq:S_latt2}
\end{equation}
where the numbers again symbolically denote all necessary indices defining the fermions' states. 
We will refer to the system described by this action as ``original system''. 
Again, we split the action into two parts $\cal{S}={\cal S}_{\mathrm{imp}}+{\cal S}_{\mathrm{rem}}$, which we will still be referring to as ``impurity'' and ``remainder'', but we wish to make clear at this point that the term ``impurity'' is used here purely for convenience, it has nothing to do with the Anderson impurity model. We also make no restrictions neither for the interaction nor for the kinetic part in terms of locality. In this sense the ``impurity'' in general spans over the whole lattice, and we do not specify whether or not the ``impurity'' model can be represented as a set of equivalent local models.
\begin{align}
{\cal S}_{\mathrm{imp}} &= -\sum_{1,2} c^*_1[G_{0,\mathrm{imp}}^{-1}]_{12}c_2+\frac14\sum_{1,2,3,4}U_{1234}c^*_1c^*_2c_3c_4, 
\label{eq:S_ref}\\
{\cal S}_{\mathrm{rem}} &= -\sum_{1,2} c^*_1[G_0^{-1}-G_{0,\mathrm{imp}}^{-1}]_{12}c_2.
\end{align}
Our goal will be to construct a diagrammatic perturbation series for the self-energy of the original model, assuming that we know everything about the reference system described by ${\cal S}_{\mathrm{imp}}$ (we will continue referring to it as ``impurity model'' for the sake of continuity), i.e. its Green's function $g$ and all $2n$-point vertices $\gamma^{(2n)}$.
We wish to show that the dual technique (and consequently the dual action) stems from the conventional weak-coupling diagrammatic technique for the original model. 
To this aim we introduce 
\begin{equation}
G'=G-g,
\end{equation}
and consider all possible skeleton diagrams for the self-energy $\Sigma$ in the conventional weak-coupling diagrammatic technique for the original system. 
We remind the reader that skeleton diagrams are diagrams that are constructed from the full rather than bare Green's functions and thus do not include diagrams that have a subgraph connected to the rest of the diagram only by two fermionic lines. 
Such diagrams are already included in the full $G$ line so we must remove them in order to prevent double-counting.
Now, instead of any diagram that has $M$ fermionic lines we consider $2^M$ expressions that are obtained from the expression for the diagram by replacing each $G$ with either $g$ or $G'$. 
We will refer to the original diagram made of $G$ lines as \emph{parent diagram}. 
As the interaction in the original lattice Hubbard model and the impurity model is identical, the topological structure of the diagrams for $\Sigma$ and $\Sigma_{\mathrm{imp}}$ is the same and differs only by the value of the fermionic lines. 
In the former it is $G$ and in the latter it is $g$. 
This in turn means that $\Sigma'=\Sigma-\Sigma_{\mathrm{imp}}$ is given the sum of all topologically distinct skeleton diagrams built from fermionic lines and bare vertices $U$, where each fermionic line is either $G'$ or $g$, but there is at least one $G'$ line.

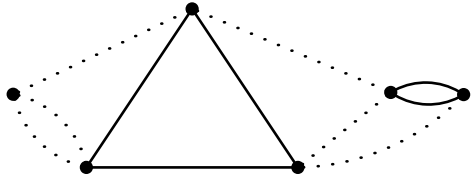
\begin{figure}
    \centering
    \begin{fmfgraph*}(200,60)
    \fmfstraight
    \fmfleft{i1}
    \fmfright{o1}
    \fmftopn{t}{6}
    \fmfbottomn{b}{6}
    \fmf{phantom, tension=10}{i1,i}
    \fmf{phantom, tension=10}{o,o1}
    \fmfdot{i}
     \fmfdot{t3}
      \fmfdot{b2}
       \fmfdot{b4}
        \fmfdot{o}
         \fmfdot{v}
    \fmf{dots,left=.3}{i,b2,i}
      \fmf{dots}{i,t3}
      \fmf{plain}{t3,b2,b4,t3}
      \fmf{dots}{t3,v}
      \fmf{dots}{b4,v}
       \fmf{dots,right=.2}{b4,o}
         \fmf{plain,left=.3,tension=2}{v,o,v}
    \end{fmfgraph*}
    \caption{A diagram giving contribution to a graph that is depicted on Fig.\ref{g_reduc_sigma}(b). In particular the bare vertex on the left contributes to a 4 point vertex. The $G'$ lines are hereafter depicted with dotted lines. The bare $U$ vertices are shown as dots rather than squares with four distinct corners in order not to overload the figure with details.}
    \label{cluster_example}
\end{figure}

The next step is to consider connected clusters in the above-mentioned diagrams that consist of bare vertices connected by $g$ lines. 
It is easy to see that those clusters contribute to the $\gamma^{(2n)}$ vertices of the impurity model. 
Indeed, due to parity reasons (each bare $U$ vertex has 4 in/out-going lines and each $g$ line has 2 ends), the number of in/out-going lines for the cluster is even, whereby we have to count not only the $G'$ lines, but also the two lines that go in and out of the parent self-energy diagram. 
On the other hand the number of those lines can not be 2, as it would violate the skeleton nature of the parent diagram. 
Also because we consider skeleton diagrams, clusters can not have loose ends (parts that are connected to the rest of the cluster with only one $g$ line and have only one in/out-going line), meaning that clusters on Fig.~\ref{gamma_reduc} are possible, but if we connect two corners of any outside square with a $g$ line it becomes forbidden, just as in the case of the impurity problem $\gamma^{(2n)}$ vertices for which such loose ends are amputated. 
The above observations mean that we can consider every diagram for $\Sigma'$ as a skeleton diagram consisting of polygons with 4, 6, 8 etc. edges connected by $G'$ lines; the polygons are contributions to the $\gamma^{(2n)}$ vertices of the impurity problem. 
A schematic example is shown on Fig.~\ref{cluster_example}.

The topology of the diagrams for $\Sigma'$ that we just obtained is identical to that of the dual diagrams, that also are nothing other than skeleton diagrams built from fermionic lines connecting polygons with even number of vertices. 
But there are several differences that we have to cope with to demonstrate how the dual technique emerges from the original weak-coupling technique.
\begin{enumerate}
    \item As we already discussed $\tilde\Sigma$ is not equal to $\Sigma'$. This is accounted for by the denominator in Eq.~\eqref{sigma_prime_dual} that diagrammatically means that $\Sigma'$ is the $g$-irreducible part of $\tilde\Sigma$.
    \item The fermionic lines in the diagrams for $\Sigma'$ and $\tilde\Sigma$ are not the same, the connection between $G'$ and $\tilde G$ has to be established.
    \item The least trivial feature is that the polygons with 6 and more vertices in the diagrams for $\Sigma'$ are though quite similar to the vertices of the impurity model, but are not identical to them. Moreover, they even do not have a universal value, meaning that a particular contribution to a polygon may or may not be present depending on which particular contributions are considered for other polygons.
\end{enumerate}

\begin{figure}
\centering
(a)    \begin{fmfgraph*}(120,80)
    \fmfleft{i}
    \fmfright{o}
    \fmfpolyn{empty,tension=0.3}{K}{6}
    \fmfpolyn{empty,tension=0.3}{L}{6}
    \fmf{phantom}{i,K1}
    \fmf{phantom}{o,L4}
    \fmf{dots,right=.4,tension=0}{K2,L3}
    \fmf{dots,tension=0.1}{K3,L2}
    \fmf{dots,tension=0.1}{K4,L1}
    \fmf{dots,tension=0.1}{K5,L6}
    \fmf{dots,left=.4,tension=0}{K6,L5}
    \end{fmfgraph*}
    (b) \begin{fmfgraph*}(120,80)
    \fmfleft{i}
    \fmfright{o}
    \fmfpolyn{empty,tension=0.3}{K}{4}
    \fmfpolyn{empty,tension=0.18}{M}{4}
    \fmfpolyn{empty,tension=0.3}{L}{6}
    \fmf{phantom}{i,K1}
    \fmf{phantom}{o,L4}
    \fmf{plain,tension=2}{K3,M1}
    \fmf{dots,right=.4,tension=0}{K2,L3}
    \fmf{dots,tension=0.1}{M2,L2}
    \fmf{dots,tension=0.1}{M3,L1}
    \fmf{dots,tension=0.1}{M4,L6}
    \fmf{dots,left=.4,tension=0}{K4,L5}
    \end{fmfgraph*}
    
    (c)
    \begin{fmfgraph*}(120,80)
    \fmfleft{i}
    \fmfright{o}
    \fmfpolyn{empty,tension=0.3}{K}{6}
    \fmfpolyn{empty,tension=0.18}{M}{4}
    \fmfpolyn{empty,tension=0.3}{L}{4}
    \fmf{phantom}{i,K1}
    \fmf{phantom}{o,L3}
    \fmf{plain,tension=2}{M3,L1}
    \fmf{dots,right=.4,tension=0}{K2,L2}
    \fmf{dots,tension=0.1}{K3,M2}
    \fmf{dots,tension=0.1}{K4,M1}
    \fmf{dots,tension=0.1}{K5,M4}
    \fmf{dots,left=.4,tension=0}{K6,L4}
    \end{fmfgraph*}
    (d)
    \begin{fmfgraph*}(120,80)
    \fmfleft{i}
    \fmfright{o}
    \fmfpolyn{empty,tension=0.3}{K}{4}
    \fmfpolyn{empty,tension=0.18}{M}{4}
    \fmfpolyn{empty,tension=0.18}{N}{4}
    \fmfpolyn{empty,tension=0.3}{L}{4}
    \fmf{phantom}{i,K1}
    \fmf{phantom}{o,L3}
    \fmf{plain,tension=2}{N3,L1}
    \fmf{plain,tension=2}{K3,M1}
    \fmf{dots,right=.4,tension=0}{K2,L2}
    \fmf{dots,tension=0.1}{M2,N2}
    \fmf{dots,tension=0.1}{M3,N1}
    \fmf{dots,tension=0.1}{M4,N4}
    \fmf{dots,left=.4,tension=0}{K4,L4}
    \end{fmfgraph*}
 \caption{Graphs (b),(c), and (d) show what graph (a) turns into if one or both hexagons are replaced by two connected squares. Graphs (b) and (c) are allowed, while the graph (d) corresponds to a non-skeleton parent diagram and is thus forbidden.}
   \label{6-6diag}
\end{figure}
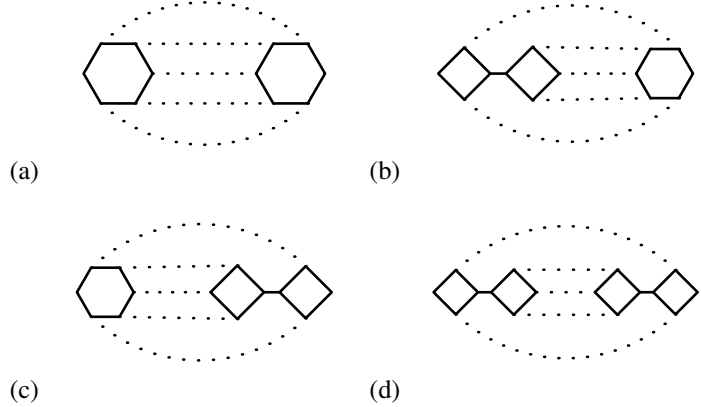

To clarify the last point let us consider the diagram on Fig.~\ref{6-6diag}\,(a). 
As we already mentioned there is a $g$-reducible contribution to $\gamma^{(6)}$ that is shown on the Fig.~\ref{gamma_reduc}\,(a). 
If we look at what happens with the diagram when we substitute one or two of the hexagons with two squares, connected by a $g$ line, we obtain graphs that are shown on the Fig.~\ref{6-6diag}\,(b-d). 
It is obvious that the last graph is illegal and does not contribute to $\Sigma'$ as the corresponding parent diagram is non-skeleton. 
Thus the $g$-reducible contribution to the hexagon is allowed if it appears only in one of two hexagons, but is not allowed if it appears in both of them.
Thus, the diagrammatics built with $G'$ lines is not related to any particular action as no universal value can be attributed to the interaction vertices.
On the other hand, if we look at the dual diagram with the same topology as on Fig.~\ref{6-6diag}\,(a) and construct the corresponding parent diagrams by depicting the $\gamma^{(6)}$ vertices as impurity diagrams, we immediately realize that parent diagram that are non-skeleton emerge, simply because in the dual technique the values of the vertices are universal and are given by the full reducible vertices of the impurity problem, including all possible $g$-reducible contributions.

Now it becomes clear how to cope with the discrepancies 1)-3) between diagrammatic representation of $\tilde\Sigma$ and $\Sigma'$. 
We have to consider all the non-skeleton parent diagrams of the dual technique, and redefine the fermionic line by absorbing those non-skeleton parts in the new fermionic line, which will eventually turn out to be $G'$. 
Of course for the sake of logical consistency it would be preferable to go the other way around --- split $G'$ into $\tilde G$ and other contributions, that, after absorbing into the polygons turn them into full impurity problem vertices. 
This is doable, but much less transparent and looks too artificial to be considered as ``diagrammatic'' derivation. 
In other words, it is hardly possible to come up with such a derivation without knowing an analytic expression connecting $G'$ with $\tilde G$. 
So we choose an aesthetically more appealing path of moving from $\tilde\Sigma$ to $\Sigma'$.

Let us consider all possible parent diagrams for the $g$-irreducible part of the (skeleton) dual diagrams. 
Now let us take any such diagram and choose some maximal dressed fermionic line (MDFL) in it. 
By MDFL we understand a subgraph connected to the rest of the diagram only at two points, that is not part of a longer subgraph of this type. 
We explicitly exclude a single $g$ line from this definition. 
It is clear that such an object must be of the form $\cal{G}\sigma\cal{G}\dots\sigma\cal{G}$, where $\cal{G}$ is a fermionic line, \emph{i.e.} either $g$ or $\tilde G$, and $\sigma$ is some subgraph with two external ends that is irreducible with respect to a single fermionic line (either $g$ or $\tilde G$). 
It is easy to see that maximum one of the $\cal{G}$ lines can be a dual propagator $\tilde G$. 
Otherwise the corresponding dual diagram would be a non-skeleton one. 
So the general form of a MDFL is $(g\sigma)^{1-s}(g\sigma)^{n}\tilde G^s(\sigma g)^{ms}g^{1-s}$, \emph{i.e.} $n$ $g$ lines separated by some $\sigma$, then $s$ $\tilde G$ lines, and finally further $ms$ $g$ lines, here $n$ and $m$ are natural numbers or zero and $s$ is 0 or 1. 
We multiply $m$ with $s$ to indicate that if the MDFL consists only of $g$ lines, then there is no need to consider the second group of $g$ lines, and the terms with $1-s$ in the exponent ensure that if the dual line is absent we start the row with $g\sigma g$. 
Now let us freeze all the rest of the considered diagram except for the MDFL and sum up all the diagrams that differ from each other only by the concrete realization of the MDFL. 
It is obvious that the result would be the frozen part of the diagram with the following object inserted in place of the MDFL:
\begin{equation}
G''=
   \begin{fmfgraph*}(20,10)
   \fmfstraight
   \fmfleft{i,d1}
   \fmfright{o,d2}
   \fmf{dashes,dash_len=.2w}{i,o}
   \end{fmfgraph*}+
   \begin{fmfgraph*}(30,10)
   \fmfstraight
   \fmfleft{i,d1}
   \fmfright{o,d2}
   \fmf{plain}{i,v}
   \fmf{plain}{v,o}
   \fmfv{d.sh=circle,d.f=empty,d.si=.4w,l.d=0,l=$\tilde\Sigma$}{v}
   \end{fmfgraph*}+
   \begin{fmfgraph*}(30,10)
   \fmfstraight
   \fmfleft{i,d1}
   \fmfright{o,d2}
   \fmf{plain}{i,v}
   \fmf{dashes,dash_len=.15w,tension=.7}{v,o}
   \fmfv{d.sh=circle,d.f=empty,d.si=.4w,l.d=0,l=$\tilde\Sigma$}{v}
   \end{fmfgraph*}+
   \begin{fmfgraph*}(30,10)
   \fmfstraight
   \fmfleft{i,d1}
   \fmfright{o,d2}
   \fmf{plain}{v,o}
   \fmf{dashes,dash_len=.15w,tension=.7}{i,v}
   \fmfv{d.sh=circle,d.f=empty,d.si=.4w,l.d=0,l=$\tilde\Sigma$}{v}
   \end{fmfgraph*}+
  \begin{fmfgraph*}(60,10)
   \fmfstraight
   \fmfleft{i,d1}
   \fmfright{o,d2}
   \fmf{plain}{i,v1}
   \fmf{dashes,dash_len=.075w,tension=.55}{v1,v2}
   \fmf{plain}{v2,o}
   \fmfvn{d.sh=circle,d.f=empty,d.si=.2w,l.d=0,l=$\tilde\Sigma$}{v}{2}
   \end{fmfgraph*}
    \label{g_prime_dual}.
\end{equation}
Or in the analytic form: $ G''=\tilde G+g\tilde\Sigma g+g\tilde\Sigma\tilde G+\tilde G\tilde\Sigma g+g\tilde\Sigma\tilde G\tilde\Sigma g$.
To understand it, it is enough to recall that $\tilde\Sigma$ is a $g$-reducible object, thus if there is anything left (or right) of $\tilde G$, it sums up to $(g\tilde\Sigma$ (or $\tilde\Sigma g$). 
Similarly, all contributions with no dual line at all sum up to $g\tilde\Sigma g$. 
Next we can do the same trick with all MDFL's of all diagrams to end up with a set of skeleton diagrams built from bare $U$ vertices and fermionic lines that are either $g$ or $G''$. 
Now we just have to persuade ourselves that $G''$ and $G'$ are identical, and this would conclude our derivation as this is precisely how $\Sigma'$ is defined, meaning $\Sigma'$ is the $g$-irreducible part of $\tilde\Sigma$.

The identity of $G'$ and $G''$ can be proven by inspection. 
We use~\eqref{sigma_prime_dual}, the definitions of $\Sigma'$ and $\tilde\Sigma$, and the expression
\begin{equation}
    \tilde G_0=\left(g^{-1}+G_0^{-1}-G_{0,\mathrm{imp}}^{-1}\right)^{-1}-g,
    \label{g_dual_def_gen}
\end{equation}
which is a trivial generalization of~\eqref{g_dual_def}. We apply following transformations:
\begin{align}
    G^{-1} &= G_0^{-1}-\Sigma = (g^{-1}+G_0^{-1}-G_{0,\mathrm{imp}}^{-1})-\Sigma' = (\tilde G_0+g)^{-1}-\Sigma' \\
    &= \left((\tilde G^{-1}+\tilde\Sigma)^{-1}+g\right)^{-1} - \tilde\Sigma(1+g\tilde\Sigma)^{-1}
    = \left(g+\tilde G+g\tilde\Sigma g+g\tilde\Sigma\tilde G+\tilde G\tilde\Sigma g+g\tilde\Sigma\tilde G\tilde\Sigma g\right)^{-1}, \notag
\end{align}
which together with the definition of $G'$ proves the identity. 
As a consequence, from the strictly diagrammatic point of view we have shown that the infinite diagrammatic row $\tilde\Sigma=\Sigma'+\Sigma'g\Sigma'+\Sigma'g\Sigma'g\Sigma'+\dots{}=\Sigma'(1-g\Sigma')^{-1}$ represents a self-energy of a theory with the interaction given by Eq.~\eqref{interaction}, and a full Green's function $\tilde G$ satisfying $ G'=\tilde G+g\tilde\Sigma g+g\tilde\Sigma\tilde G+\tilde G\tilde\Sigma g+g\tilde\Sigma\tilde G\tilde\Sigma g.$ For the sake of completeness the last expression can be inverted to yield 
\begin{equation}
    \tilde G=G'-g\Sigma'g-g\Sigma'G'-G'\Sigma'g+g\Sigma'G'\Sigma'g+g\Sigma'g\Sigma'g.
    \label{g_dual_full}
\end{equation}
Then, knowing the self-energy and the full Green's function of the theory, we can calculate the bare Green's function which of course happens to be the well-known $\tilde G_0$ given by (\ref{g_dual_def_gen}). This completes the diagrammatic derivation of the dual theory from the conventional weak-coupling diagrammatics.

The highly non-trivial part of this derivation is that $G'$ happens to be identical with $G''$. The consequence of this fact is that we not just merely reshuffled the diagrams of the original technique, but also constructed a new action. The building blocks of that new action, i.e. the propagator~\eqref{g_dual_full} and the interaction~\eqref{interaction} emerge from this reshuffling of the diagrams, which is not something one would \emph{a priori} expect.

\section{Truncation of the dual fermion scheme}

The proposed scheme in its complete form is clearly unsuitable for practical applications. As mentioned above, going beyond the $\gamma^{(4)}$ vertex in the dual interaction is computationally extremely difficult, not to mention considering the whole infinite row of interaction vertices. So for any practical implementations of the scheme a truncation is necessary. Strictly speaking any truncation would destroy the consistency of the theory in the sense that one will no longer be able to identify the MDFL's with $G'$. The natural question that arises is to which degree the relation~\eqref{sigma_prime_dual} is affected if instead of the exact value of $\tilde\Sigma$ we take some approximation. In this section we argue that this relation should be kept unaltered, \emph{i.e.} the correction to $\Sigma_{\rm imp}$ stemming from the dual theory is non-additive.  

An important lesson of the diagrammatic derivation presented above is that the full lattice Green's function $G$ of the original problem~\eqref{eq:S_latt2} cannot be easily separated into the reference (``impurity'') and remainder contributions that are respectively given by $g$ and $\tilde{G}$ Green's functions.
The total Green's function can still be expressed as $G = g + G'$.
However, as we explicitly showed above, it is impossible to rigorously construct a theory using $G'$ lines as building blocks for a diagrammatic expansion.
As a consequence, the total self-energy also cannot be simply obtained as a sum of corresponding contributions $\Sigma_{\rm imp}$ and $\tilde{\Sigma}$, contrary to a common belief of diagrammatic theories formulated on the basis of the DMFT impurity reference system~\cite{Sun02, Biermann03, Ayral12, Ayral13, Huang14, Boehnke16, Ayral17, Ayral15, Ayral16, Ayral17-2}.
This results in a more complicated relation for the lattice self-energy~\eqref{sigma_prime_dual}. 
As discussed above, the denominator $(1+g\tilde{\Sigma})$ in this expression isolates only those contributions to $\tilde{\Sigma}$ that are irreducible with respect to the impurity Green's function $g$. Now, if we consider only $\gamma^{(4)}$ vertices in the dual diagrammatics one might naively assume that as those vertices are $g$-irreducible, the denominator in~\eqref{sigma_prime_dual} is redundant. However, the role of this denominator can also be viewed from a different angle that outlines its importance even in the case where an approximation for $\tilde\Sigma$ is $g$-irreducible.

To illustrate this point, we first partially dress the bare Green's function $G_{0}$ of the original model~\eqref{eq:S_latt2} with the ``impurity'' self-energy via the Dyson equation. This results in the DMFT-like Green's function $G^{\rm DMFT}=(G_0^{-1}-\Sigma_{\mathrm{imp}})^{-1}$ and leads to the following expression for the full Green's function
\begin{align}
G = G^{\rm DMFT} + G^{\rm DMFT} T G^{\rm DMFT}.
\end{align}
Here, we additionally introduced a $T$-matrix that satisfies
\begin{align}
T^{-1} = \Sigma'^{-1} - G^{\rm DMFT},
\label{eq:T_matrix1}
\end{align}
which can be equivalently rewritten as 
\begin{align}
T^{-1} = \tilde{\Sigma}^{-1} - \tilde{G}_{0}
\label{eq:T_matrix2}
\end{align}
using Eq.~\eqref{sigma_prime_dual} and the fact that $G^{\rm DMFT} = \tilde{G}_{0} + g$.
Therefore, one can see that in the diagrammatic expression for the $T$-matrix, the self-energies $\tilde{\Sigma}$, which represent corrections to the impurity self-energy, can be connected between themselves only by the dual Green's function $\tilde{G}$ and not by the impurity Green's function $g$.
The reason is that the self-energy $\tilde{\Sigma}$ is related to the dual theory~\eqref{eq:S_dual}, where building blocks for the diagrammatic expansion contain only dual Green's functions and impurity vertices. Therefore, in the exact theory impurity vertices cannot be connected by the impurity line $g$, since the latter is not present in the dual problem~\eqref{eq:S_dual}.
This has been pointed out already in the Ref.~\cite{DUPUIS2001617}.
The dual self-energy $\tilde{\Sigma}$ fulfills this requirement by construction. 
Thus, the possibility to connect impurity vertices by the impurity Green's function has to be excluded only from the Dyson equation for the $T$-matrix~\eqref{eq:T_matrix1}. 
This exclusion is performed exactly with the help of the denominator $(1+g\tilde{\Sigma})$ that enters the Eq.~\eqref{sigma_prime_dual}, which leads to the final diagrammatic form of the $T$-matrix~\eqref{eq:T_matrix2}. Should one omit the denominator and merely replace $\Sigma'$ with $\tilde\Sigma$ in~\eqref{eq:T_matrix1}, one would generate contributions to the $T$-matrix that include impurity vertices connected by impurity Green's functions as the latter are part of $G^{\rm DMFT}$, in contradiction with the general property of the exact theory. We view this as an important indication that, contrary to the findings of~\cite{Katanin_2013} using the Eq.~\eqref{sigma_prime_dual} is crucial for obtaining reasonable results with the truncated dual scheme, which is also supported by numerical data~\cite{Rohringer_RMP_2018, DiagDFQMC}.

\section{General reference system: Plaquette example.}

\begin{figure}[t!]
 \centering
\includegraphics[width=0.65\textwidth]{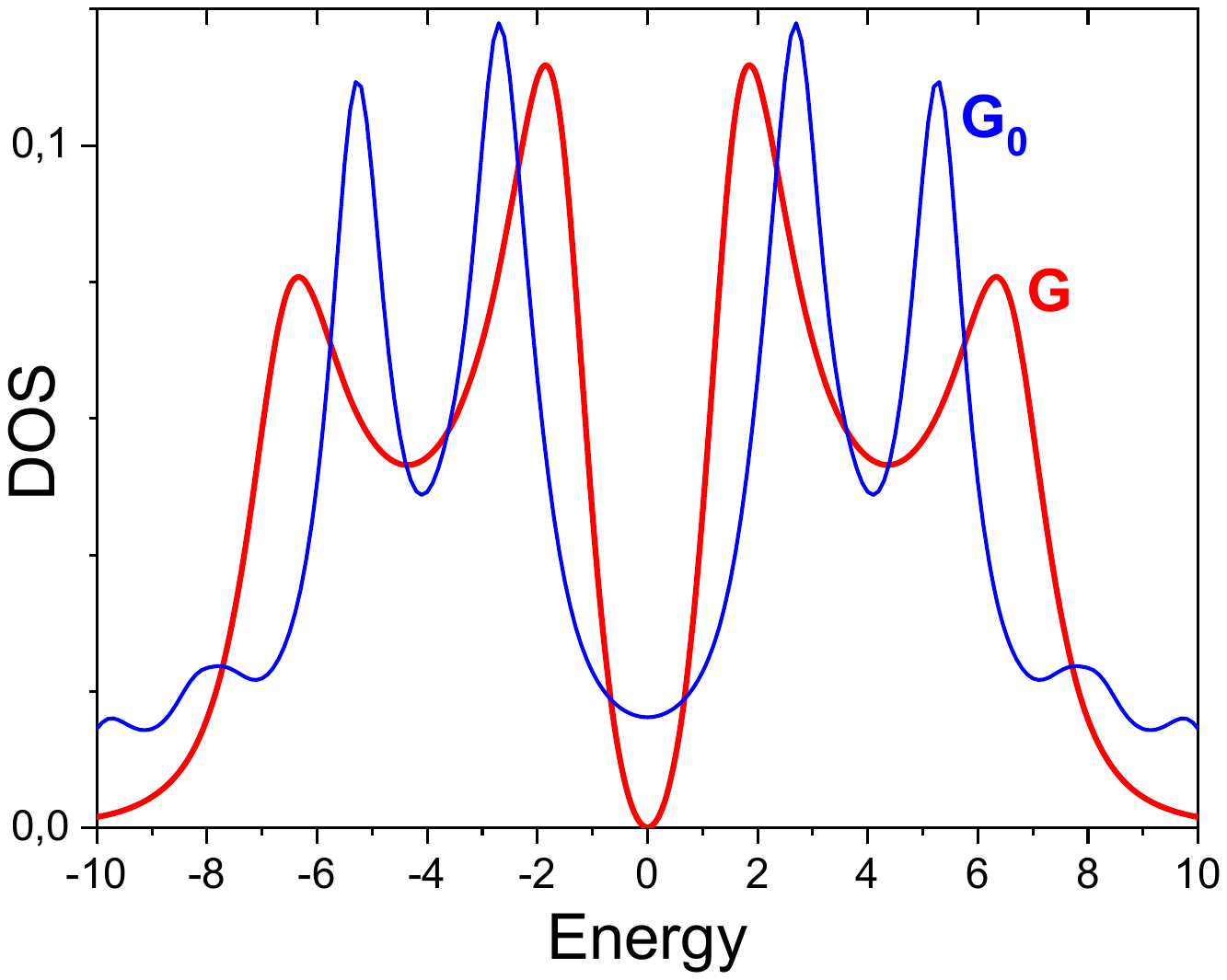}
 \caption{
Density of states for dual fermion second-order perturbation from plaquette for $U=W=8$ ($G$)
in comparison with the DOS for the reference (plaquette) system ($G_0$).
}
\label{DOS}
\end{figure}

As mentioned earlier the dual fermion technique has the reputation of ``beyond DMFT'' extension. We wish to once again draw the reader's attention to the fact that becomes particularly clear due to the diagrammatic derivation above, that the reference system for its construction is in no way restricted to the Anderson impurity model.

For a simple numerical example we choose a $2\times2$ plaquette as a reference system~\eqref{eq:S_ref} to describe the half-filled two-dimension Hubbard model~\eqref{eq:S_latt}. 
We choose parameters for the strong coupling regime  $U=W=8t$ and $T=0.3t$.
In the standard cluster DMFT scheme~\cite{DMFT, CDMFT} the bare Green's function $G_{0, {\rm imp}}$ (or equivalently the hybridization function) of the reference system is determined from the self-consistency condition~\eqref{eq:condition} for the lattice Green's function $G({\bf k})$ averaged over a supercell Brillouin zone.
Here, we explore another possibility and chose the bare Green's function of the reference system to be equivalent to the bare Green's function of the initial model at ${\bf k}=0$, namely $G_{0, {\rm imp}} = G_{0}({\bf k}=0)$.
In other words, we define the hybridization function of a $2\times2$ plaquette as the $4\times4$ matrix of hoppings taken at zero momentum in momentum space representation $\Delta=t_{{\bf k} = 0}$, which corresponds to a lattice of decoupled plaquettes with periodic boundary conditions.       
Note that the single-particle spectrum of such periodic plaquette coincides with the tight-binding energies at 
the $2\times2$ $\mathbf{k}$-grid in original Brillouin zone.
In this sense, we can view the dual fermion perturbation from the plaquette reference system as a multi-grid interpolation in the $\mathbf{k}$-space. It is different from simple cluster size extrapolations in the DCA approach~\cite{Maier_DCA,Maier_DCApp} or nested cluster approximation in CDMFT scheme~\cite{Parcollet_CDMFT,Parcollet_NCDMFT}. 

With this choice of the reference system, one can use the exact diagonalization scheme to calculate the dual Green function and the plaquette vertex function~\cite{Hafermann_2009}.
The Fig.~\ref{DOS} shows the density of states (DOS) for  the reference plaquette and the result for the second-order plaquette dual-fermion~\cite{PhysRevB.79.045133}. 
We use Pad\'e-analytical continuation from Matsubara to the real energy axes~\cite{DMFT}. 
We conclude that the DOS for the dual fermion theory differs from original reference system and is related to the renormalization of the low-energy singlet peaks in the plaquette towards the Fermi level to form ``Slatter-peaks'' in the lattice, as well as moving the high-energy ``Hubbard-peaks'' in the opposite direction\cite{Gull_2008,Harland2020}.

\begin{figure}[t!]
 \centering
\includegraphics[width=0.49\textwidth,angle=0]{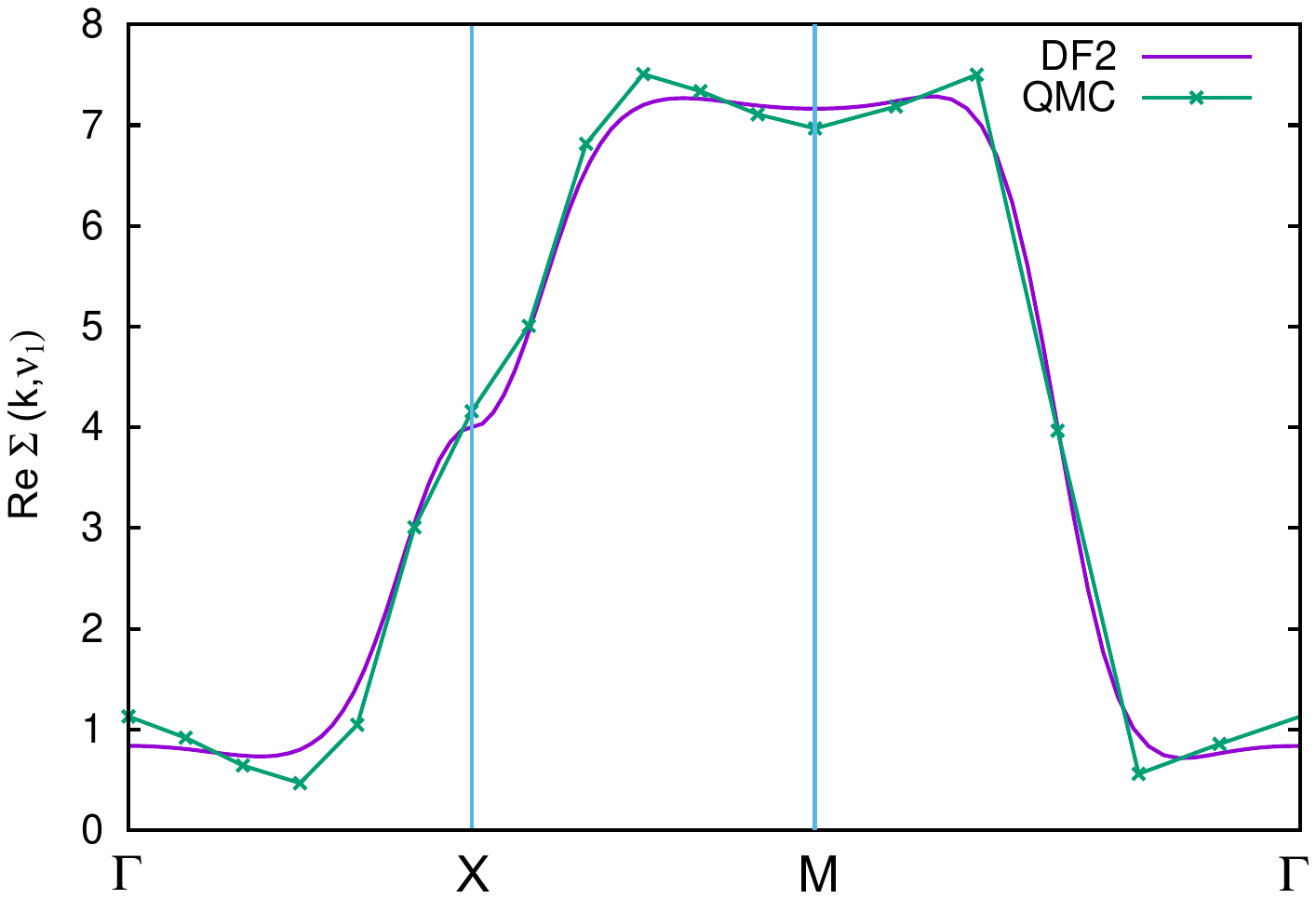}
\includegraphics[width=0.49\textwidth,angle=0]{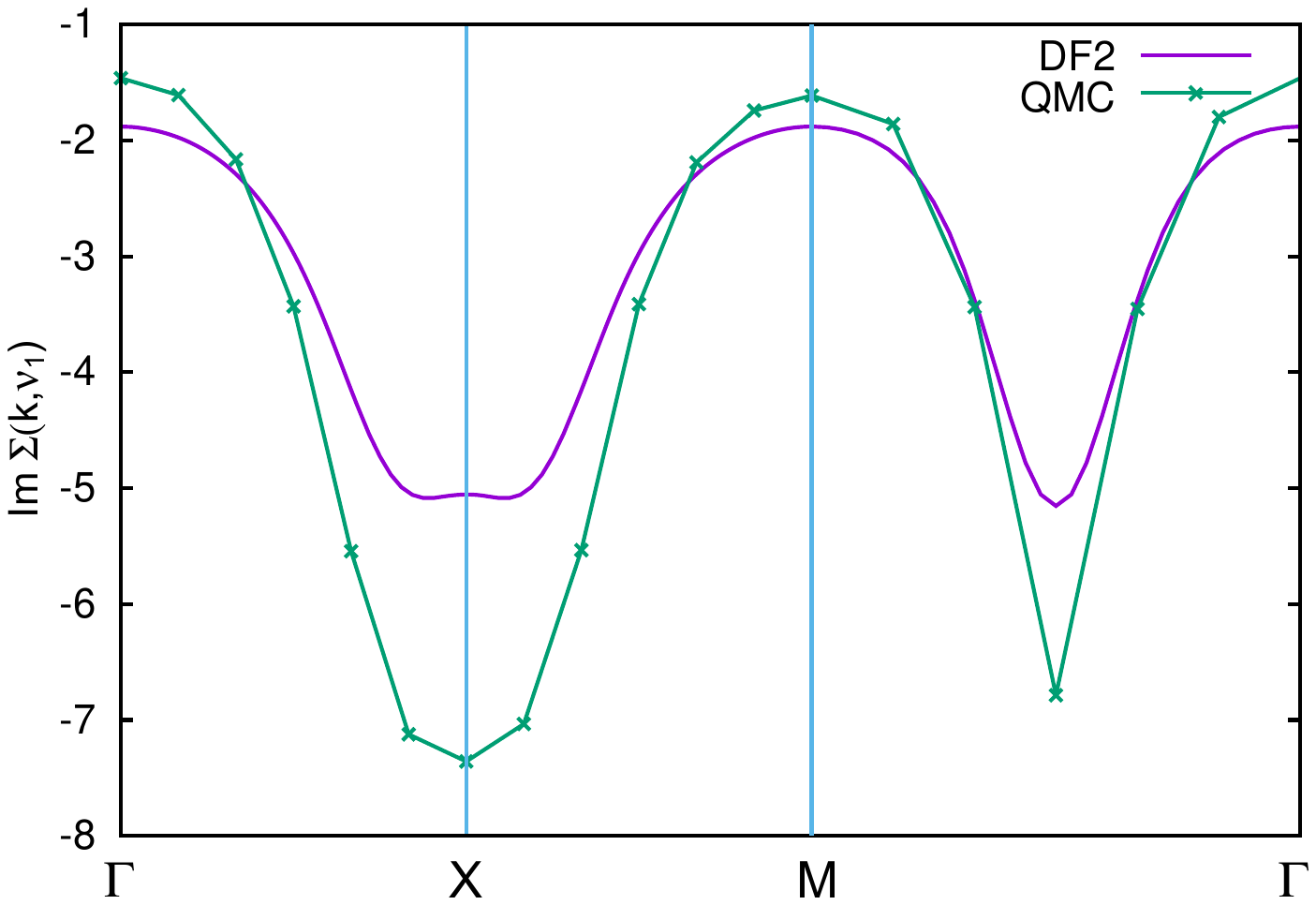}
 \caption{
Real (left) and imaginary (right) part of the self-energy for the DF plaquette scheme in
comparison with diagrammatic-QMC results\cite{DiagDFQMC}
at the first Matsubara frequency}
\label{Sig}
\end{figure}

The Fig.~\ref{Sig} shows the dual fermion plaquette lattice self-energy in the second order aproximation for the standard ${\mathbf{k}}$-dependent high-symmetry path $\Gamma-X-M-\Gamma$ in the two-dimensional Brillouin zone, together with numerically exact lattice diagrammatic QMC~\cite{DiagDFQMC}.
The almost perfect agreement for real part of $\Sigma({\bf k}, \nu=\pi T)$ shows the strength of the dual fermion 
superperturbation technique starting from a reasonable plaquette reference system. We expect more elaborate approaches like ladder dual fermion to significantly improve the results. A more thorough development of this path will be presented in an upcoming publication.

\section{Conclusions and outlook}
We have provided a diagrammatic dirivation of the dual fermion formalism starting from the conventional weak-coupling diagrammatics. The derivation is based the skeleton diagram approach and basically consists of splitting the fermionic lines of the original system into the ``impurity'' line and the residual line. Upon this the contributions are regrouped to single out $n$-particle vertices of the impurity problems as the elements of the dual technique. Finally it is shown that the obtained diagrammatics corresponds to the dual action with the vertices of the ``impurity'' problem serving as its bare interaction.

We find that the residual self-energy is given by the impurity line-irreducible part of the dual self-energy, and discuss the nature of the reducibility of the dual self-energy. We argue that this property is essential and that the self-energy of the original problem is intrinsically non-additive, \emph{i.e.} not equal to the sum of the ``impurity'' and dual self-energies.

We stress that the derivation does not rely on any particular form of the reference ``impurity'' system. In particular we show a simple test with the reference system for the Hubbard model being a set of disconnected $2\times2$ plaquettes with periodic boundary conditions. Already the lowest orders of the dual self-energy provide very satisfactory results.

We also expect this approach to be helpful for other problems, e.g. one might choose a particle-hole symmetric Hubbard model as the reference model and account for finite next-nearest-neighbour hopping and finite doping perturbatively. Also this approach could be useful outside of condensed matter theory if a problem can be significantly simplified by modifying the quadratic part of the action.

\section*{Acknowledgments}
The authors thank Andrei Katanin for useful comments.
S.B and A.I.L acknowledge support by the Cluster of Excellence 'Advanced Imaging of Matter' of the Deutsche Forschungsgemeinschaft (DFG) - EXC 2056 - project ID 390715994 and by the DFG SFB 925.
The work of M.I.K. and A.I.L. is supported by European Research Council via Synergy Grant 854843 - FASTCORR.

\bibliographystyle{elsarticle-num}
\bibliography{Ref.bib}

\end{fmffile}

\end{document}